# Robust Dynamic State Estimator of Integrated Energy Systems based on Natural Gas Partial Differential Equations

Liang Chen, *Member*, *IEEE*, Yang Li, *Senior Member IEEE*, Manyun Huang, Xinxin Hui, and Songlin Gu

***Abstract*—The reliability and precision of dynamic database are vital for the optimal operating and global control of integrated energy systems. One of the effective ways to obtain the accurate states is state estimations. A novel robust dynamic state estimation methodology for integrated natural gas and electric power systems is proposed based on Kalman filter. To take full advantage of measurement redundancies and predictions for enhancing the estimating accuracy, the dynamic state estimation model coupling gas and power systems by gas turbine units is established. The exponential smoothing technique and gas physical model are integrated in Kalman filter. Additionally, the time-varying scalar matrix is proposed to conquer bad data in Kalman filter algorithm. The proposed method is applied to an integrated gas and power systems formed by GasLib-40 and IEEE 39-bus system with five gas turbine units. The simulating results show that the method can obtain the accurate dynamic states under three different measurement error conditions, and the filtering performance are better than separate estimation methods. Additionally, the proposed method is robust when the measurements experience bad data.**

## I. Introduction

THE penetration of large-scale renewable energies increases the random fluctuation of power supplies, making power systems more vulnerable [1]. Aiming at this problem, the concept of power system flexibility is proposed [2], [3]. The gas turbine units (GTUs) can respond to the power fluctuation of renewable energies more rapidly than the coal-fired power units [4], providing a technical means for enhancing the power system flexibility [5], [6]. The global coordinating optimization strategy to enhance the flexibility of integrated natural gas and electric power systems (IGESs) attracts many researchers' attentions [7], [8].

L. Chen is with the School of Automation, Nanjing University of Information Science and Technology, Nanjing 210044, China (e-mail: ch.lg@nuist.edu.cn).

Y. Li is with the School of Electrical Engineering, Northeast Electric Power University, Jilin 132012, China (e-mail: liyang@neepu.edu.cn).

M.Y. Huang is with the College of Energy and Electrical Engineering, Hohai University, Nanjing, 210096, China (e-mail: hmy_hhu@yeah.net).

X.X. Hui is with the Weifang Power Supply Company State Grid Shandong Electric Power Company, Weifang 261061, China (e-mail: 1578716468@qq.com).

S.L. Gu is with the State Grid Economic and Technological Research Institute CO.,LTD. Beijing, 210008, China (e-mail: gsl0516@163.com).

The accurate real-time dynamic database is essential for the advanced applications of IGESs, such as optimal control of generations [9] and optimal energy flows [10]. Although some important points of IGESs are equipped with measuring devices, it is impossible to capture the entire useful states by measurements. On the other hand, the measuring devices experience random errors and big deviations due to transmission errors or interferences inevitably [11], [12], the data acquired from which cannot be applied directly. State estimations are effective ways to filter random errors of measurements and establish an accurate database for real time applications under the dynamic process of IGESs.

Weighted least squares (WLSs) based static state estimations (SSEs) [13] filter random errors depending on the measurement redundancy, which is applicable for static operating conditions of both natural gas and power systems. However, WLSs estimate the states only based on measurement redundancies, omitting the dynamic information of time domains. Aiming at this problem, Kalman filter based forecasting-aided state estimations of power systems [14] predict the states one step ahead using exponential smoothing techniques [15] [16], which is one of most popular extrapolation algorithm used in estimation problems because of the good overall performance and relative simplicity of computations. For natural gas pipeline networks, the dynamic processes of pressures and mass flow rates of gases caused by the continuous changes of gas loads and supplies can be presented as a set of nonlinear partial differential equations (PDEs) derived from the momentum conservation principles and materiel balances [17]-[20]. By using the PDEs to predict the gas states, the Kalman filter based dynamic state estimations (DSEs) for gas pipeline networks are proposed in [21]-[23]. In [21], the DSE model of gas system is built. In [22], the PDEs are discretized by finite element methods. In [23], a joint state and parameter estimation problem is considered. However, these methods estimate natural gas states only, which are not DSEs of IGESs.

The forecasting-aided state estimations of power systems and DSE based on PDEs of gas systems can be integrated to form the DSE methodology for IGESs. In practice, both of the scale and measurement redundancy of gas systems are smaller than power systems, and the gas measurements are always used for billing requirements rather than state estimations. The DSE of IGESs can take full advantage of power system measurements to enhance the estimating accuracy of gas systems by the coupling constraints. On the other hand, the PDEs of gas systems provide additional constraints for power systems by coupling points. Therefore, these two systems should be coupled to estimate the states and the global optimized estimation results can be obtained compared with the separate estimation methods. Although state estimations of interconnected electric and gas networks have already been presented in [24][25], essentially, they are SSEs rather than a DSE because the time domain constraints of gas PDEs and power system quasi statics are not considered.

The existing filtering algorithms, just like unscented Kalman filter (UKF) and cubature Kalman filter (CKF) can deal with non-linear problems. However, UKF is not applicable to large scale system, and the matrix in CKF may be singular for IGESs. As a result, the nonlinear PDEs of gas systems have to be linearized in Kalman filter algorithms. In [26][27], the dynamic states are redefined as deviations from steady states for linearization purpose. Actually, the pressures and flows are changing all the time during dynamic processes in IGESs rather than steady states. In addition, the following challenges are faced: 1) the power and gas states are predicted by exponential smoothing and physical models of gases, respectively. The exponential smoothing can be presented as a difference equation, while the physical models are PDEs. The two kinds of mathematical equations have to be integrated in Kalman filter. 2) The coupling constraints of these two systems in Kalman filter have to be modeled. 3) The boundary conditions are not differential equations, making Kalman filter hard to be modeled and solved.

In our previous work [28], we propose a Kalman filter based DSE methodology for IGESs. The simplified DSE model of IGESs that ignoring compressors is established. The coupling model of the energy transformation between gas and electric power systems through GTUs is built. The boundary conditions are taken as supplementary constraints of the PDEs in Kalman filter. A simple integrated energy system is constructed to study the method. In this paper, both of the models, algorithm and simulations are enhanced. The more accurate DSE model of IGESs considering compressors and virtual nodes in natural pipeline networks is built. Additionally, to enhance the robustness of the algorithm, the time-varying scalar matrix is proposed to conquer the bad data in measurements. The proposed DSE method is applied to a standard IGES formed by GasLib-40 and IEEE 39-bus system with five gas turbine units, and the results of the proposed integrated DSEs are compared with the results of the separated estimation methods. The simulation testifies the advantages of the proposed method.

The rest of this paper is organized as follows. Section II builds the dynamic model of power systems using the Holt's exponential smoothing techniques. Section III establishes the linearized system model of gas pipeline networks. Section IV gives the DSE methods of IGESs based on the Kalman filter. To evaluate the performance of DSE method quantitatively, the case study is carried out under different noise conditions in Section V. Finally, Section VI concludes this paper.

## II. Dynamic Model of Power Systems

Due to the good overall performance and relative simplicity of computations, Holt's exponential smoothing is used to predict power system states:

$$\boldsymbol{L}_t = \alpha_\mathrm{S}\boldsymbol{x}_t^\mathrm{E} + (1-\alpha_\mathrm{S})(\boldsymbol{L}_{t-1}+\boldsymbol{T}_{t-1}) \tag{1}$$

$$\boldsymbol{T}_t = \beta_\mathrm{S}(\boldsymbol{L}_t-\boldsymbol{L}_{t-1}) + (1-\beta_\mathrm{S})\boldsymbol{T}_{t-1} \tag{2}$$

$$\boldsymbol{x}_{t+1}^\mathrm{E} = \boldsymbol{L}_t + \boldsymbol{T}_t \tag{3}$$

where, $\boldsymbol{x}_t^\mathrm{E} = [e_{1t}\ f_{1t}\ e_{2t}\ f_{2t} \cdots e_{n_\mathrm{B}t}\ f_{n_\mathrm{B}t}]^\mathrm{T}$ is the power system state vector, $\boldsymbol{x}_t^\mathrm{E} \in \mathbb{R}^{2n_\mathrm{B}\times 1}$, $n_\mathrm{B}$ is the number of buses; $e_{it}$ and $f_{it}$ are the bus voltage real and imaginary parts at time $t$, respectively; $0<\alpha_\mathrm{S}, \beta_\mathrm{S}<1$ are the smoothing parameters. $\boldsymbol{L}_t$ is the length of the average for the estimation of the level; $\boldsymbol{T}_t$ is the smoothing of the trend. The optimal values of $\alpha_\mathrm{S}$ and $\beta_\mathrm{S}$ are obtained by minimizing the sum of squared errors of the one-step-ahead predictions. The initial values of $\boldsymbol{L}_t$ and $\boldsymbol{T}_t$ are $\boldsymbol{x}_2^\mathrm{E}$ and $\boldsymbol{x}_2^\mathrm{E} - \boldsymbol{x}_1^\mathrm{E}$, respectively. Equations (1) and (2) are substituted into (3), and then (1)-(3) can be written as the following matrix form:

$$\boldsymbol{x}_{t+1}^\mathrm{E} = \alpha_\mathrm{S}\boldsymbol{x}_t^\mathrm{E} + \boldsymbol{u}_{t+1}^\mathrm{E} \tag{4}$$

where, $\boldsymbol{u}_{t+1}^\mathrm{E} = (1-\alpha_\mathrm{S})(\boldsymbol{L}_{t-1}+\boldsymbol{T}_{t-1}) + \boldsymbol{T}_t$.

With the rapid development of phasor measurement units (PMUs), the PMU measurements can be taken as the DSE measurement vector. The advantage of PMUs is that phasors can be measured directly, and the measurement equations are linear, which facilitates the modeling and solving of Kalman filter. The PMU measurements include bus voltage, branch current and bus injected current phasors. The measurement vector of power system at time instant $t$ is $\boldsymbol{z}_t^\mathrm{E} = \left[\left(\boldsymbol{z}_t^\mathrm{V}\right)^\mathrm{T}\ \left(\boldsymbol{z}_t^\mathrm{IB}\right)^\mathrm{T}\ \left(\boldsymbol{z}_t^\mathrm{IN}\right)^\mathrm{T}\right]^\mathrm{T}$, in which $\boldsymbol{z}_t^\mathrm{V} = [\cdots\ e_{b,t}^\mathrm{Z}\ f_{b,t}^\mathrm{Z}\ \cdots]^\mathrm{T}$, $\boldsymbol{z}_t^\mathrm{V} \in \mathbb{R}^{2n_\mathrm{ZB}\times 1}$, $\boldsymbol{z}_t^\mathrm{IB} = [\cdots\ I_{ij,t}^\mathrm{BR}\ I_{ij,t}^\mathrm{BI}\ \cdots]^\mathrm{T}$, $\boldsymbol{z}_t^\mathrm{IB} \in \mathbb{R}^{2n_\mathrm{ZC}\times 1}$, $\boldsymbol{z}_t^\mathrm{IN} = [\cdots\ I_{b,t}^\mathrm{IR}\ I_{b,t}^\mathrm{II}\ \cdots]^\mathrm{T}$, $\boldsymbol{z}_t^\mathrm{IN} \in \mathbb{R}^{2n_\mathrm{ZB}\times 1}$. $e_{b,t}^\mathrm{Z}$ and $f_{b,t}^\mathrm{Z}$ are the measurements of voltage real and imaginary parts at bus $b$, respectively; $I_{ij,t}^\mathrm{BR}$ and $I_{ij,t}^\mathrm{BI}$ are the current real and imaginary part measurements of branch $ij$, respectively; $n_\mathrm{ZB}$ is the number of buses equipped with PMUs; $I_{b,t}^\mathrm{IR}$ and $I_{b,t}^\mathrm{II}$ are the injected current real and imaginary part measurements of bus $b$; $n_\mathrm{ZC}$ is the number of branch current measurements, respectively. The equation between $\boldsymbol{z}_t^\mathrm{E}$ and $\boldsymbol{x}_t^\mathrm{E}$ is

$$\boldsymbol{z}_t^\mathrm{E} = \begin{bmatrix}\boldsymbol{z}_t^\mathrm{V}\\ \boldsymbol{z}_t^\mathrm{IB}\\ \boldsymbol{z}_t^\mathrm{IN}\end{bmatrix} = \begin{bmatrix}\boldsymbol{H}_\mathrm{V}\\ \boldsymbol{H}_\mathrm{IB}\\ \boldsymbol{H}_\mathrm{IN}\end{bmatrix}\boldsymbol{x}_t^\mathrm{E} = \boldsymbol{H}_\mathrm{E}\boldsymbol{x}_t^\mathrm{E} \tag{5}$$

The details of $\boldsymbol{H}_\mathrm{V}$, $\boldsymbol{H}_\mathrm{IB}$ and $\boldsymbol{H}_\mathrm{IN}$ are shown in Appendix A.

## III. Model of Gas Pipeline Networks

### A. Partial Differential Equations of Gas Flow in Pipelines

The gas dynamics in large scale pipeline networks cannot be ignored. The fluid dynamic process of compressible gas along horizontal pipelines can be presented as a set of PDEs as follows [17][23]:

$$\frac{\partial \rho}{\partial \tau} + \frac{\partial(\rho v_\mathrm{G})}{\partial \varsigma} = 0 \tag{6}$$

$$\frac{\partial(\rho v_\mathrm{G})}{\partial \tau} + \frac{\partial(\rho v_\mathrm{G}^2)}{\partial \varsigma} + \frac{\partial p}{\partial \varsigma} = -\frac{\gamma \rho v_\mathrm{G}\,|\,v_\mathrm{G}\,|}{2d} \tag{7}$$

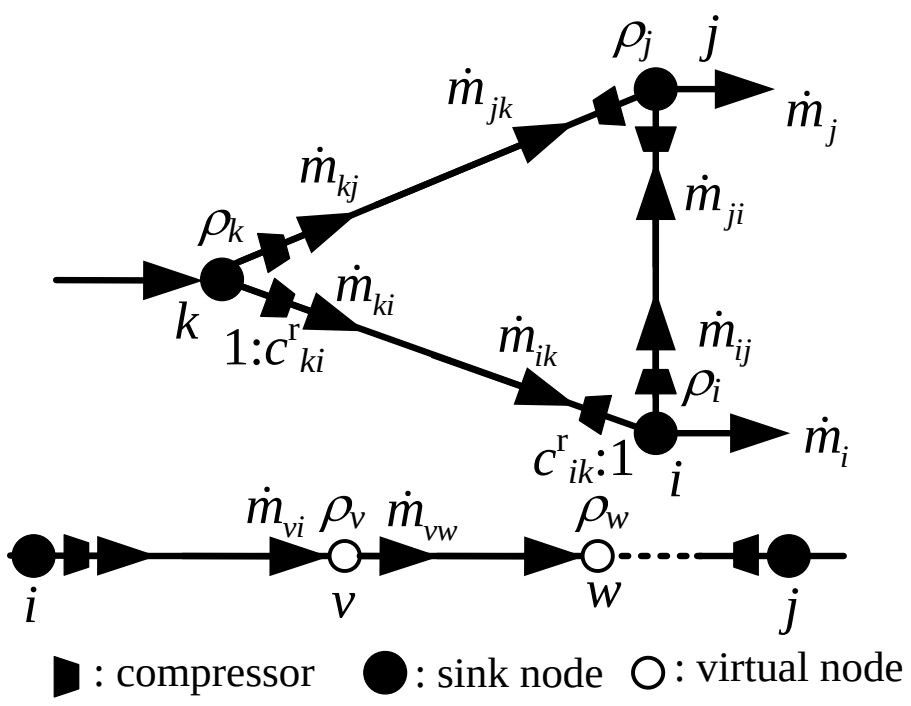

Fig. 1. The model of natural gas pipeline networks.

where, $\rho$ and $p$ are the gas density and pressure; $v_{\mathrm{G}}$ is the gas velocity flowing along pipelines; $\tau$ and $\varsigma$ are the time and space along pipelines, respectively; $d$ and $\gamma$ are the pipeline diameter and the friction factor between natural gases and pipeline inner surfaces, respectively. Equations (6) and (7) are the mass and momentum balance equations. Practically, the gas velocity in pipelines is much smaller than the sound speed, as a result, the second term of (7), describing the convective effect of the natural gas, can be omitted. The relationships between the pressure $p$, mass flow rate of gases in pipelines $\dot{m}$ and density $\rho$ are as follows [17]:

$$p = c_{\mathrm{S}}^2 \rho \tag{8}$$

$$\dot{m} = \rho v_{\mathrm{G}} a \tag{9}$$

where, $c_{\mathrm{S}}$ is the constant sound speed; $a$ is the cross section area of pipelines. The PDE (7) is nonlinear, however, equations in Kalman filter are linear, so (7) should be linearized. Aiming at this problem, $|v_{\mathrm{G}}|$ is replaced by the absolute value of a constant average gas velocity $|\bar{v}_{\mathrm{G}}|$, thus (6) and (7) become

$$\frac{\partial \rho}{\partial \tau} + \frac{\partial \dot{m}}{a \partial \varsigma} = 0 \tag{10}$$

$$\frac{\partial \dot{m}}{a \partial \tau} + c_{\mathrm{S}} \frac{\partial \rho}{\partial \varsigma} + \frac{\gamma \dot{m} |\bar{v}_{\mathrm{G}}|}{2da} = 0 \tag{11}$$

The model of pipeline networks is shown in Fig. 1. The node numbers $i$, $j$ and $k$ satisfy $k < i < j$, and the gas flow direction is from the small number node to the big one. To ensure the model accuracy, the long pipelines should be separated into several segments. The nodes connecting two segments are virtual nodes, which should be treated as a node in the pipeline model. The mass flow rate going out from the sink node $i$ and virtual node $v$ are $\dot{m}_i$ and 0, respectively. For simplification, the compressor at $k$ end of pipeline $ki$ is modeled as a constant ratio of densities $c^{\mathrm{r}}_{ki}$. The linear PDEs (10) and (11) are solved by the Euler finite difference technique:

$$\begin{aligned} & c^{\mathrm{r}}_{ji}\rho_{j,t+1} - c^{\mathrm{r}}_{ji}\rho_{j,t} + c^{\mathrm{r}}_{ij}\rho_{i,t+1} - c^{\mathrm{r}}_{ij}\rho_{i,t} \\ & + \frac{\Delta t}{L^{\mathrm{P}}_{ij} a_{ij}} \left( \dot{m}_{ji,t+1} - \dot{m}_{ij,t+1} + \dot{m}_{ji,t} - \dot{m}_{ij,t} \right) = 0 \end{aligned} \tag{12}$$

$$\begin{aligned} & \dot{m}_{ji,t+1} - \dot{m}_{ij,t+1} + \dot{m}_{ji,t} - \dot{m}_{ij,t} \\ & + \frac{a_{ij}\Delta t c^2}{L^{\mathrm{P}}_{ij}} \left( c^{\mathrm{r}}_{ji}\rho_{j,t+1} - c^{\mathrm{r}}_{ij}\rho_{i,t+1} + c^{\mathrm{r}}_{ji}\rho_{j,t} - c^{\mathrm{r}}_{ij}\rho_{i,t} \right) \\ & + \frac{\gamma |\bar{v}_{\mathrm{G}}| \Delta t}{4 d_{ij} a_{ij}} \left( \dot{m}_{ij,t+1} + \dot{m}_{ji,t+1} + \dot{m}_{ij,t} + \dot{m}_{ji,t} \right) = 0 \end{aligned} \tag{13}$$

where, the subscripts $i$, $ij$ and $t$ represent node $i$, pipeline $ij$, and time $t$, respectively; $L^{\mathrm{P}}_{ij}$ is the length of pipeline $ij$; $\Delta t$ is the time step; $\dot{m}_{ij,t}$ is the mass flow rate from $i$ to $j$ in pipeline $ij$.

### B. Boundary Conditions

The PDEs are less than the gas states, therefore, in addition to the PDEs, the gas dynamics should satisfy some algebraic equations named boundary conditions. The unique solution could be solved if and only if both of the PDEs and boundary conditions are given. In this paper, the gas nodes are classified into three types: source nodes, sink nodes and virtual nodes. The source node connects a gas station, and the gas density of the source node is assumed to be constant. The sink node connects a gas load, or several pipelines. The sink node cannot store gases, so the gas injection and outflow at a sink node should be equal, i.e. the material balance constraint. The virtual nodes connect two segments, so the mass flow rates at the two sides of virtual node $v$ are equal, shown in Fig.1. The constant gas density at source nodes, material balance constraint at sink nodes and the equal mass flow rates at the two sides of virtual nodes are the boundary conditions, represented as follows:

$$\rho_{s,t} = \rho^{\mathrm{C}}_s, \quad s\text{: source node} \tag{14}$$

$$\sum_k \dot{m}_{ik,t} - \sum_j \dot{m}_{ij,t} = \dot{m}_{i,t}, \; k, j \in i, \; k < i, \; j > i, \; i\text{: sink node} \tag{15}$$

$$\dot{m}_{vi,t} - \dot{m}_{vw,t} = 0, \; v\text{: virtual node} \tag{16}$$

where, $k, j \in i$ represents node $k$ and $j$ are connected to $i$; $\rho^{\mathrm{C}}_s$ is the constant gas density at node $s$; $\dot{m}_{i,t}$ is the mass flow rate going out from node $i$ at time $t$.

The PDEs and boundary conditions constitute the system model of natural gas pipeline networks, the number of which is equal to the gas states.

### C. System Model

The gas densities and mass flow rates at the two ends of pipelines are taken as states. For a $n_{\mathrm{N}}$-node and $n_{\mathrm{P}}$-pipeline gas network, the number of states is $n_{\mathrm{N}} + 2n_{\mathrm{P}}$. The state vector at time $t$ is $\boldsymbol{x}^{\mathrm{G}}_t = \left[ \left(\boldsymbol{x}^{\mathrm{Gr}}_t\right)^{\mathrm{T}} \; \left(\boldsymbol{x}^{\mathrm{Gm}}_t\right)^{\mathrm{T}} \right]^{\mathrm{T}}$, $\boldsymbol{x}^{\mathrm{Gr}}_t = [\rho_{1,t}, \rho_{2,t}, \cdots, \rho_{n_{\mathrm{N}},t}]^{\mathrm{T}}$, $\boldsymbol{x}^{\mathrm{Gm}}_t = [\cdots, \dot{m}_{ij,t}, \dot{m}_{ji,t}, \cdots]^{\mathrm{T}}$. $\boldsymbol{x}^{\mathrm{Gr}}_t \in \mathbb{R}^{n_{\mathrm{N}} \times 1}$, $\boldsymbol{x}^{\mathrm{Gm}}_t \in \mathbb{R}^{2n_{\mathrm{P}} \times 1}$. Equations (12) and (13) can be rewritten in the following matrix form:

$$\begin{bmatrix} \boldsymbol{A}_{11} & \boldsymbol{A}_{12} \\ \boldsymbol{A}_{21} & \boldsymbol{A}_{22} \end{bmatrix} \begin{bmatrix} \boldsymbol{x}^{\mathrm{Gr}}_{t+1} \\ \boldsymbol{x}^{\mathrm{Gm}}_{t+1} \end{bmatrix} = \begin{bmatrix} \boldsymbol{A}_{11} & -\boldsymbol{A}_{12} \\ -\boldsymbol{A}_{21} & -\boldsymbol{A}_{22} \end{bmatrix} \begin{bmatrix} \boldsymbol{x}^{\mathrm{Gr}}_t \\ \boldsymbol{x}^{\mathrm{Gm}}_t \end{bmatrix} \tag{17}$$

The details of $\boldsymbol{A}_{11}$, $\boldsymbol{A}_{12}$, $\boldsymbol{A}_{21}$ and $\boldsymbol{A}_{22}$ are shown in Appendix B. The matrix form of boundary conditions (14)-(16) is

$$\begin{bmatrix} \boldsymbol{B}_{11} & \boldsymbol{0}_{n_{\mathrm{S}}\times 2n_{\mathrm{P}}} \\ \boldsymbol{0}_{n_{\mathrm{SI}}\times n_{\mathrm{N}}} & \boldsymbol{B}_{22} \\ \boldsymbol{0}_{n_{\mathrm{V}}\times n_{\mathrm{N}}} & \boldsymbol{B}_{33} \end{bmatrix} \begin{bmatrix} \boldsymbol{x}_t^{\mathrm{Gr}} \\ \boldsymbol{x}_t^{\mathrm{Gm}} \end{bmatrix} = \begin{bmatrix} \boldsymbol{u}_t^{\mathrm{r}} \\ \boldsymbol{u}_t^{\mathrm{m}} \\ \boldsymbol{0}_{n_{\mathrm{V}}\times 1} \end{bmatrix} \tag{18}$$

where, $\boldsymbol{u}_t^{\mathrm{r}} = [\ldots, \rho_s^{\mathrm{C}}, \ldots]^{\mathrm{T}}$, $s$ is source node, $\boldsymbol{u}_t^{\mathrm{r}} \in \mathbb{R}^{n_{\mathrm{S}}}$; $\boldsymbol{u}_t^{\mathrm{m}} = [\ldots, \dot{m}_{i,t}, \ldots]^{\mathrm{T}}$, $i$ is sink node, $\boldsymbol{u}_t^{\mathrm{m}} \in \mathbb{R}^{n_{\mathrm{SI}}}$; $n_{\mathrm{S}}$, $n_{\mathrm{SI}}$ and $n_{\mathrm{V}}$ are the numbers of source nodes, sink nodes and virtual nodes, respectively; $\boldsymbol{0}$ is the zero matrix, the subscript of which is the dimension. The details of $\boldsymbol{B}_{11}$, $\boldsymbol{B}_{22}$ and $\boldsymbol{B}_{33}$ are shown in Appendix C.

Equations (17) and (18) can be written as

$$\mathcal{A}\boldsymbol{x}_{t+1}^{\mathrm{G}} = \mathcal{B}\boldsymbol{x}_t^{\mathrm{G}} + \mathcal{U}_{t+1} \tag{19}$$

$$\mathcal{A} = \begin{bmatrix} \boldsymbol{A}_{11} & \boldsymbol{A}_{12} \\ \boldsymbol{A}_{21} & \boldsymbol{A}_{22} \\ \boldsymbol{B}_{11} & \boldsymbol{0}_{n_{\mathrm{S}}\times 2n_{\mathrm{P}}} \\ \boldsymbol{0}_{n_{\mathrm{SI}}\times n_{\mathrm{N}}} & \boldsymbol{B}_{22} \\ \boldsymbol{0}_{n_{\mathrm{V}}\times n_{\mathrm{N}}} & \boldsymbol{B}_{33} \end{bmatrix}, \mathcal{B} = \begin{bmatrix} \boldsymbol{A}_{11} & -\boldsymbol{A}_{12} \\ -\boldsymbol{A}_{21} & -\boldsymbol{A}_{22} \\ \boldsymbol{0}_{n_{\mathrm{S}}\times n_{\mathrm{N}}} & \boldsymbol{0}_{n_{\mathrm{S}}\times 2n_{\mathrm{P}}} \\ \boldsymbol{0}_{n_{\mathrm{SI}}\times n_{\mathrm{N}}} & \boldsymbol{0}_{n_{\mathrm{SI}}\times 2n_{\mathrm{P}}} \\ \boldsymbol{0}_{n_{\mathrm{V}}\times n_{\mathrm{N}}} & \boldsymbol{0}_{n_{\mathrm{V}}\times 2n_{\mathrm{P}}} \end{bmatrix}, \mathcal{U}_{t+1} = \begin{bmatrix} \boldsymbol{0}_{n_{\mathrm{P}}\times 1} \\ \boldsymbol{0}_{n_{\mathrm{P}}\times 1} \\ \boldsymbol{u}_{t+1}^{\mathrm{r}} \\ \boldsymbol{u}_{t+1}^{\mathrm{m}} \\ \boldsymbol{0}_{n_{\mathrm{V}}\times 1} \end{bmatrix}.$$

The system model of pipeline networks is obtained

$$\boldsymbol{x}_{t+1}^{\mathrm{G}} = \boldsymbol{F}_{\mathrm{G}}\boldsymbol{x}_t^{\mathrm{G}} + \boldsymbol{u}_{t+1}^{\mathrm{G}} \tag{20}$$

where, $\boldsymbol{F}_{\mathrm{G}} = \mathcal{A}^{-1}\mathcal{B}$, $\boldsymbol{u}_{t+1}^{\mathrm{G}} = \mathcal{A}^{-1}\mathcal{U}_{t+1}$.

### D. Measurement Equation

The node pressures and mass flows of gas load can be measured if meters are equipped at the nodes, so the gas measurement vector includes pressures and load flows. The measurement equation is

$$\boldsymbol{z}_{t+1}^{\mathrm{G}} = \boldsymbol{H}_{\mathrm{G}}\boldsymbol{x}_{t+1}^{\mathrm{G}} \tag{21}$$

where, $\boldsymbol{z}_{t+1}^{\mathrm{G}} = [z_{1,t+1}^{\mathrm{r}}, z_{2,t+1}^{\mathrm{r}}, \ldots, z_{n_{\mathrm{ZP}},t+1}^{\mathrm{r}}, z_{1,t+1}^{\mathrm{m}}, z_{2,t+1}^{\mathrm{m}}, \ldots, z_{n_{\mathrm{ZF}},t+1}^{\mathrm{m}}]^{\mathrm{T}}$; $z_{i,t+1}^{\mathrm{r}}$ and $z_{i,t+1}^{\mathrm{m}}$ are the pressure and load flow measurements at time $t$+1, respectively. The details of $\boldsymbol{H}_{\mathrm{G}}$ are shown in Appendix D.

## IV. Robust Dynamic State Estimation of IGESs

### A. Coupling Model of IGESs

In IGESs, the natural gas pipeline networks and power system are coupled by the GTUs, by which the electric power is produced from natural gases. The efficiency is considered to be constant in this paper. The relationship between gas flow and power is presented as:

$$P_{i,t}^{\mathrm{G}} = \eta_i \dot{m}_{s,t},\ i \prec s \tag{22}$$

where, $P_{i,t}^{\mathrm{G}}$ and $\eta_i$ are the output power at time $t$ and energy conversion coefficient of the GTU connected to bus $i$; $i \prec s$ represents that the gas source of the GTU connected to bus $i$ is the sink node $s$. In power system, $P_{i,t}^{\mathrm{G}}$ satisfy the bus injected power equation, which is shown as follows:

$$P_{i,t}^{\mathrm{G}} = \sum_j [e_{it}(G_{ij}^{\mathrm{Y}} e_{jt} - B_{ij}^{\mathrm{Y}} f_{jt}) + f_{it}(G_{ij}^{\mathrm{Y}} f_{jt} + B_{ij}^{\mathrm{Y}} e_{jt})],\ j \mapsto i \tag{23}$$

where, $G_{ij}^{\mathrm{Y}}$ and $B_{ij}^{\mathrm{Y}}$ are the real and imaginary parts of the $i^{\mathrm{th}}$ row $j^{\mathrm{th}}$ column element in power system admittance matrix, respectively; $j \mapsto i$ represents bus $j$ and bus $i$ are connected by a transmission line. The relationship between gas flow and power system states can be derived from (22) and (23):

$$\dot{m}_{s,t} = \frac{1}{\eta_i}\sum_j [e_{it}(G_{ij}^{\mathrm{Y}} e_{jt} - B_{ij}^{\mathrm{Y}} f_{jt}) + f_{it}(G_{ij}^{\mathrm{Y}} f_{jt} + B_{ij}^{\mathrm{Y}} e_{jt})] \tag{24}$$

### B. Dynamic State Estimation Model of IGESs

The state vector of IGESs, denoted as $\boldsymbol{x}_t^{\mathrm{I}}$, includes the electric power system states $\boldsymbol{x}_{\mathrm{E}t}$ and natural gas pipeline network states $\boldsymbol{x}_t^{\mathrm{G}}$, i.e. $\boldsymbol{x}_t^{\mathrm{I}} = \left[\left(\boldsymbol{x}_t^{\mathrm{E}}\right)^{\mathrm{T}}\ \left(\boldsymbol{x}_t^{\mathrm{G}}\right)^{\mathrm{T}}\right]^{\mathrm{T}}$, and the measurements $\boldsymbol{z}_t^{\mathrm{I}} = \left[\left(\boldsymbol{z}_t^{\mathrm{E}}\right)^{\mathrm{T}}\ \left(\boldsymbol{z}_t^{\mathrm{G}}\right)^{\mathrm{T}}\right]^{\mathrm{T}}$. The DSE model of IGESs includes the system equation and measurement equation, shown as follows:

$$\begin{cases} \boldsymbol{x}_{t+1}^{\mathrm{I}} = \boldsymbol{F}_{\mathrm{I}}\boldsymbol{x}_t^{\mathrm{I}} + \boldsymbol{u}_{t+1}^{\mathrm{I}} + \boldsymbol{v} \\ \boldsymbol{z}_{t+1}^{\mathrm{I}} = \boldsymbol{H}_{\mathrm{I}}\boldsymbol{x}_{t+1}^{\mathrm{I}} + \boldsymbol{w} \end{cases} \tag{25}$$

where, $\boldsymbol{F}_{\mathrm{I}} = \begin{bmatrix} \alpha_{\mathrm{S}}\boldsymbol{I}_{2n_{\mathrm{B}}\times 2n_{\mathrm{B}}} & \boldsymbol{0}_{2n_{\mathrm{B}}\times(2n_{\mathrm{P}}+n_{\mathrm{N}})} \\ \boldsymbol{0}_{2n_{\mathrm{P}}\times 2n_{\mathrm{B}}} & \boldsymbol{F}_{\mathrm{G}} \end{bmatrix}$,

$\boldsymbol{u}_{t+1}^{\mathrm{I}} = \left[\left(\boldsymbol{u}_{t+1}^{\mathrm{E}}\right)^{\mathrm{T}}\ \left(\boldsymbol{u}_{t+1}^{\mathrm{G}}\right)^{\mathrm{T}}\right]^{\mathrm{T}}$,

$\boldsymbol{H}_{\mathrm{I}} = \begin{bmatrix} \boldsymbol{H}_{\mathrm{E}} & \boldsymbol{0}_{(4n_{\mathrm{ZB}}+2n_{\mathrm{ZC}})\times(2n_{\mathrm{P}}+n_{\mathrm{N}})} \\ \boldsymbol{0}_{(n_{\mathrm{ZP}}+n_{\mathrm{ZF}})\times 2n_{\mathrm{B}}} & \boldsymbol{H}_{\mathrm{G}} \end{bmatrix}$; $\boldsymbol{v}$ and $\boldsymbol{w}$ are the predicting and measurement errors, respectively, $\boldsymbol{v} \in \mathbb{R}^{(2n_{\mathrm{B}}+n_{\mathrm{N}}+2n_{\mathrm{P}})\times 1}$, $\boldsymbol{w} \in \mathbb{R}^{(4n_{\mathrm{ZB}}+2n_{\mathrm{ZC}}+n_{\mathrm{ZP}}+n_{\mathrm{ZF}})\times 1}$. The variance matrixes of $\boldsymbol{v}$ and $\boldsymbol{w}$ are $\boldsymbol{Q}$ and $\boldsymbol{R}$, respectively. The assumption has to be made that the parameters in dynamic equations are accurate, so the values of $\boldsymbol{Q}$ are very small. The errors of different measurements are independent, therefore $\boldsymbol{R}$ is a diagonal matrix. The diagonal elements of $\boldsymbol{R}$ are the variances of measurement errors.

In (25), the state vector $\boldsymbol{x}_{t+1}^{\mathrm{I}}$ has to be predicted by system equations using the state vector $\boldsymbol{x}_t^{\mathrm{I}}$ at time $t$. However, the gas PDEs is less than the gas states, so the PDEs cannot be solved. Aiming at this problem, the algebraic equations of boundary conditions are used as supplementary constraints in system equations. All of the states in algebraic equations have to be in the same time instance. Because the time instant of predicted states is $t$+1, the variable values of boundary conditions should be at $t$+1 also. Next, the method that obtains the variable values at $t$+1 using the values at $t$ is presented. The values of boundary condition vectors $\boldsymbol{u}_{t+1}^{\mathrm{r}}$ and $\boldsymbol{u}_{t+1}^{\mathrm{m}}$ derived from $\mathcal{U}_{t+1}$ are obtained by the following method:

a. For the constant gas density at source nodes, the variable values of $\boldsymbol{u}_{t+1}^{\mathrm{r}}$ are constant.
b. The boundary condition (15) represents the mass flow balance of sink nodes, the right part of which is the mass flow of gas loads at instant $t$+1. For the gas loads that supply gases for GTUs, the variable is the supplied gas flow at time $t$+1, the value of which can be obtained by (24). The voltage $e_{it+1}$ and $f_{it+1}$ in (24) at instant $t$+1 are predicted by Holt's exponential smoothing.
c. For the other sink nodes, the variables are the mass flows of gas loads at time $t$+1, also, the value of which can be predicted by the Holt's exponential smoothing techniques directly. For the virtual nodes, the variables are 0.

### C. Robust Kalman Filter

In this work, Kalman filter is used to solve the DSE problem. For the model (25), the processes include two steps [22]:

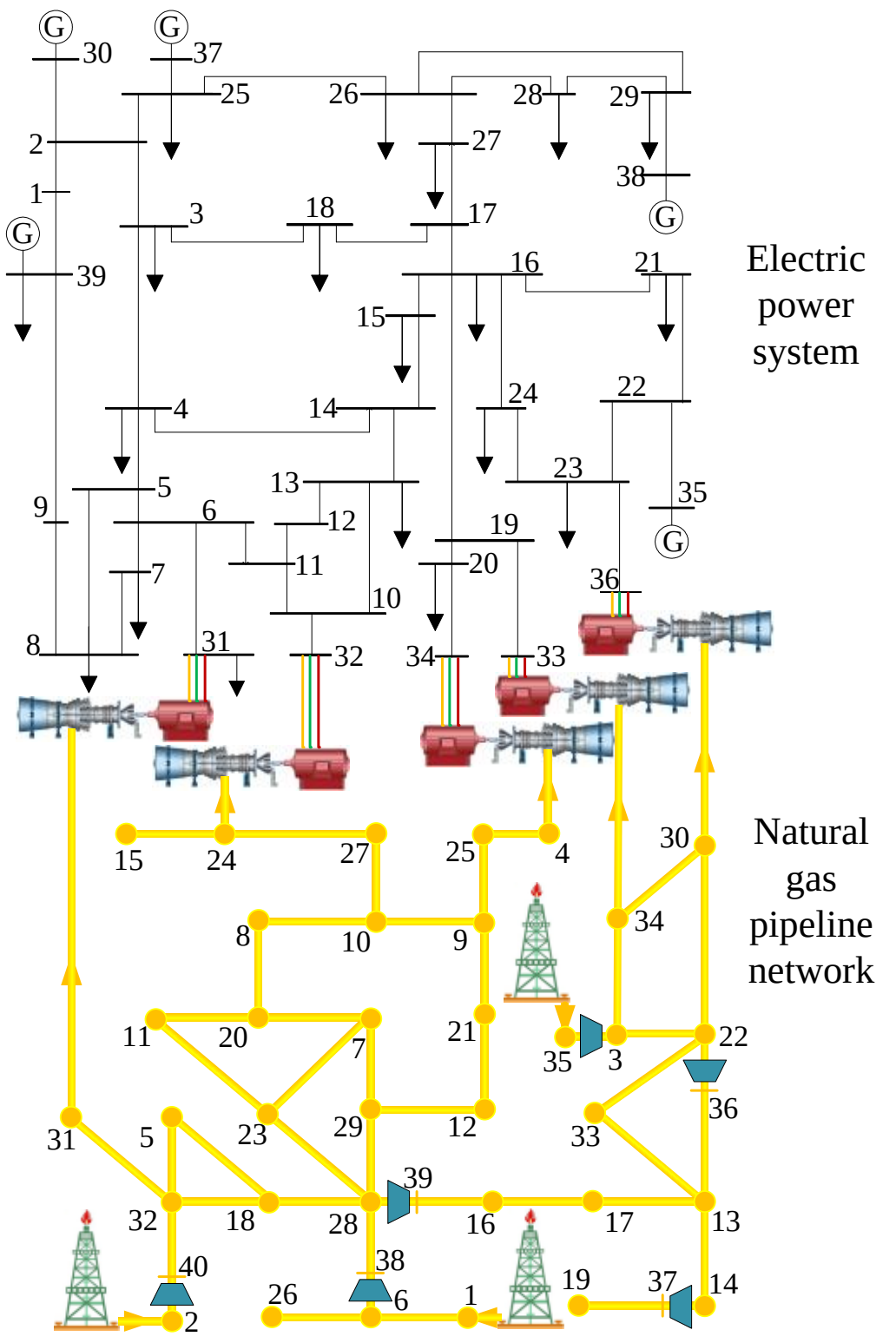

Fig. 2. The IGES testing system. The electric power system and natural gas pipeline network are coupled by five GTUs.

**Prediction step.** The predicted value of states $\tilde{\boldsymbol{x}}_{t+1}^{\mathrm{I}}$ and predicting covariance matrix $\boldsymbol{P}_{t+1|t}^{\mathrm{I}}$ are obtained according to

$$\tilde{\boldsymbol{x}}_{t+1}^{\mathrm{I}} = \boldsymbol{F}_{\mathrm{I}}\hat{\boldsymbol{x}}_{t}^{\mathrm{I}} + \boldsymbol{u}_{t+1}^{\mathrm{I}} \tag{26}$$

$$\boldsymbol{P}_{t+1|t}^{\mathrm{I}} = \boldsymbol{F}_{\mathrm{I}}\boldsymbol{P}_{t|t}^{\mathrm{I}}\boldsymbol{F}_{\mathrm{I}}^{\mathrm{T}} + \boldsymbol{Q} \tag{27}$$

**Filtering step.** The estimated state $\hat{\boldsymbol{x}}_{t+1}^{\mathrm{I}}$ and estimating covariance matrix $\boldsymbol{P}_{t+1|t+1}^{\mathrm{I}}$ are computed.

$$\boldsymbol{K}_{t+1}^{\mathrm{I}} = \boldsymbol{P}_{t+1|t}^{\mathrm{I}}\boldsymbol{H}_{\mathrm{I}}^{\mathrm{T}}(\boldsymbol{H}_{\mathrm{I}}\boldsymbol{P}_{t+1|t}^{\mathrm{I}}\boldsymbol{H}_{\mathrm{I}}^{\mathrm{T}} + \boldsymbol{R})^{-1} \tag{28}$$

$$\hat{\boldsymbol{x}}_{t+1}^{\mathrm{I}} = \tilde{\boldsymbol{x}}_{t+1}^{\mathrm{I}} + \boldsymbol{K}_{t+1}^{\mathrm{I}}(\boldsymbol{z}_{t+1}^{\mathrm{I}} - \boldsymbol{H}_{\mathrm{I}}\tilde{\boldsymbol{x}}_{t+1}^{\mathrm{I}}) \tag{29}$$

$$\boldsymbol{P}_{t+1|t+1}^{\mathrm{I}} = \boldsymbol{P}_{t+1|t}^{\mathrm{I}} - \boldsymbol{K}_{t+1}^{\mathrm{I}}\boldsymbol{H}_{\mathrm{I}}\boldsymbol{P}_{t+1|t}^{\mathrm{I}} \tag{30}$$

The estimating performance of the above Kalman filter algorithm decreases if the measurements experience bad data. To solve this problem, a time-varying scalar based robust Kalman filter is proposed to solve the dynamic state estimation problem of IGESs. The concept of innovation covariance matrix $\boldsymbol{P}_{t+1}^{\mathrm{E}}$ is defined firstly:

$$\boldsymbol{P}_{t+1}^{\mathrm{E}} = \boldsymbol{H}_{\mathrm{I}}\boldsymbol{P}_{t+1|t}^{\mathrm{I}}\boldsymbol{H}_{\mathrm{I}}^{\mathrm{T}} + \boldsymbol{\mu}_{t+1}\boldsymbol{R} \tag{31}$$

where, $\boldsymbol{\mu}_{t+1}$ is a time-varying scalar matrix. The innovation covariance matrix should satisfy the following constrain:

$$\boldsymbol{P}_{t+1}^{\mathrm{E}} = \frac{1}{m_{\mathrm{W}}}\sum_{i=0}^{m_{\mathrm{W}}-1}\boldsymbol{e}_{t+1-i}\boldsymbol{e}_{t+1-i}^{\mathrm{T}} = \boldsymbol{H}_{\mathrm{I}}\boldsymbol{P}_{t+1|t}^{\mathrm{I}}\boldsymbol{H}_{\mathrm{I}}^{\mathrm{T}} + \boldsymbol{\mu}_{t+1}\boldsymbol{R} \tag{32}$$

where, $\boldsymbol{e}_{t+1-i}$ is innovation vector, $\boldsymbol{e}_{t+1-i} = \boldsymbol{z}_{t+1-i} - \boldsymbol{H}_{\mathrm{I}}\tilde{\boldsymbol{x}}_{t+1}^{\mathrm{I}}$; $m_{\mathrm{W}}$ is the sliding window length. $\boldsymbol{\mu}_{t+1}$ can be obtained by:

$$\boldsymbol{\mu}_{t+1} = \left(\frac{1}{m_{\mathrm{w}}}\sum_{i=0}^{m_{\mathrm{w}}-1}\boldsymbol{e}_{t+1-i}\boldsymbol{e}_{t+1-i}^{\mathrm{T}} - \boldsymbol{H}_{\mathrm{I}}\boldsymbol{P}_{t+1|t}^{\mathrm{I}}\boldsymbol{H}_{\mathrm{I}}\right)\boldsymbol{R}_{t+1}^{-1} \tag{33}$$

The scalar matrix $\boldsymbol{\mu}_{t+1}$ obtained from (33) is non-diagonal, which may cause the inversion of $\boldsymbol{P}_{t+1}^{\mathrm{E}}$ singular. To solve this problem, a new diagonal matrix $\boldsymbol{\mu}_{t+1}'$ is defined:

$$\boldsymbol{\mu}_{t+1}' = \mathrm{diag}(\mu_1', \mu_2', \cdots \mu_{2n_{\mathrm{Z}}}') \tag{34}$$

where, $\mu_i' = \max(1, \mu_{t+1,ii})$; $\mu_{t+1,ii}$ is the diagonal element of $\boldsymbol{\mu}_{t+1}$; $n_{\mathrm{Z}}$ is the total number of measurements. Equation (28) becomes

$$\boldsymbol{K}_{t+1}^{\mathrm{I}} = \boldsymbol{P}_{t+1|t}^{\mathrm{I}}\boldsymbol{H}_{\mathrm{I}}^{\mathrm{T}}(\boldsymbol{H}_{\mathrm{I}}\boldsymbol{P}_{t+1|t}^{\mathrm{I}}\boldsymbol{H}_{\mathrm{I}}^{\mathrm{T}} + \boldsymbol{\mu}_{t+1}'\boldsymbol{R})^{-1} \tag{35}$$

The dynamic state estimation can be solved by the robust Kalman filter (26), (27), (35), (29) and (30). Under normal measurement condition, i.e. measurements without bad data, the diagonal elements in $\boldsymbol{\mu}_{t+1}'$ are 1, and the proposed robust Kalman filter is same to the traditional Kalman filter (26)-(30). If the measurements experience bad data, the diagonal elements values of $\boldsymbol{\mu}_{t+1}'$ increase and the algorithm shows robustness to the bad data.

## V. Case Studies

An IGES with GasLib-40 [29] natural gas system and IEEE 39-bus power system coupling by five GTUs is constructed, shown in Fig. 2. The changing processes of the electric power loads and GTUs' output power are simulated artificially. The power system states are obtained by power flow calculation using Matpower [30]. The mass flows of GTUs are calculated by (22) according to the output power. The dynamic states of pipeline networks are calculated by (20). The simulated values are true values, while the measured values are generated by adding specific distributed random numbers to the true values. The calculation step is 10 minutes, and the simulation time is 24 hours.

After the description of the testing system, the estimation is carried out under the following measurement random errors:

a. White Gaussian noises.

b. Non-zero mean noises. The measurement errors obey normal distribution, but the mean values are not zero.

c. Non-Gaussian noises.

Additionally, the bad data condition is also considered.

### A. *Description of the Testing System*

The testing IGES includes five GTUs. The energy conversion coefficient here is 20.148 MW·s/kg under 40% efficiency. The friction factor $\gamma$ and the gas speed are 0.003 and 340 m/s respectively. Gas node 1, 2 and 35 are source nodes. The pressures of these nodes are constant in the dynamic processes, and the values are 41.48 bar, 41.48 bar and 42.16 bar, respectively. The smoothing parameters $\alpha_{\mathrm{S}}$ and $\beta_{\mathrm{S}}$ are 0.5 and 0.4, respectively. It should be noted that node 35 is source node, and it connects node 3 by a compressor. So node 3 is taken as a source node, and the pressure is constant. Node 35 is deleted. Additionally, the output sides of other compressors like 36-40 are not taken as a node, so the total node number is 34.

The following buses are equipped with PMUs, and the voltages and injected currents can be measured: 2, 4, 6, 9, 12, 15, 18, 21, 22, 25, 28, 30, 31, 32, 33, 34, 35, 36, 37, 38 and 39. The following branch currents can be measured: 39-1, 25-2, 30-2, 15-3, 6-5, 7-6, 11-6, 31-6, 9-8, 39-9, 11-10, 13-10, 32-10, 14-13, 15-14, 17-16, 18-17, 27-17, 20-19, 33-19, 34-20, 35-22,

24-23, 36-23, 26-25, 37-25, 28-26, 29-26 and 38-29. The following nodes are installed with meters, and the pressures and load mass flows can be measured: 4, 5, 8, 10, 11, 15, 19, 20, 22, 24, 26, 29, 30, 31 and 34.

### B. White Gaussian Noises

In this section, the DSE is carried out under the condition that all of the measurement errors are white Gaussian noises. The standard deviations of voltage and mass flow measurement errors are 2%, respectively, while the standard deviation of pressure measurement is 0.5%-2%. The values of measurement variance matrix ***R*** are the measurement error variances, i.e. the squares of standard deviations. The values of the predictive error variance matrix ***Q*** are set to a small value $10^{-5}$ here.

To evaluate the performance of the DSE method quantitatively, the filter coefficient $\varepsilon_1$ and the total variance of estimation errors $\varepsilon_2$ are defined [31], [32]:

$$\varepsilon_1 = \sum_{t=1}^{S}(\hat{x}_t - x_t^+)^2 \Big/ \sum_{t=1}^{S}(x_t^M - x_t^+)^2 \tag{36}$$

$$\varepsilon_2 = \sum_{t=1}^{S}(\hat{x}_t - x_t^+)^2 \Big/ S \tag{37}$$

where, $\hat{x}_t$, $x_t^M$ and $x_t^+$ are the estimated value, measured value and true value, respectively; $S$ is the total sampling number.

The filter coefficients and total variances of gas pipeline states and electric power states are shown partly in Tab. I -Tab. III. It can be seen that the filter coefficients are smaller than 1, meaning that the DSE method is effective. The total variances of estimation errors are smaller than the measurement error variance.

### C. Non-zero Mean Noises

In practical systems, the errors of measuring equipments do not always satisfy the zero-mean distribution. The condition that the mean values of measurement errors are not zero should be considered. In this section, based on the measurements in section *B*, 0.5 bar deviation is added to the pressure measurement of node 24; 0.4 kg/s deviation is added to the mass flow measurement of node 4; 0.02 deviation is added to the voltage real and imaginary part measurements of bus 12. The estimating results and estimating errors are shown in Fig. 3 and Fig.4, respectively. It can be seen that the measurements deviate from the true values, but the estimated values are not influenced seriously.

The filter coefficients and total variances under non-zero mean noises are presented partly in Tab. IV. It is can be seen that the filter coefficients are still smaller than 1 even the error deviations exist. The total variances increase along with the raise of deviations. This is because the estimating results are influenced by measurement errors. However, the filter coefficients decrease, because estimation errors increase slower than measurement errors.

### D. Non-Gaussian Noises

In addition to the normal distribution, the errors of practical measuring equipments satisfy the Cauchy distribution or Laplace distribution [31], [33]. In this section, the random numbers satisfying Cauchy and Laplace distributions are added to the true values of voltages and gas pipeline states, respectively, which are taken as measurements. The central value is 0, and the scaling parameters are 0.5% - 1% for the pressures, 2% and 1% for voltages and mass flows.

TABLE I
FILTER COEFFICIENTS UNDER WHITE GAUSSIAN NOISES

| Node | Pressure ($10^{-4}$) | Mass | Node | Pressure ($10^{-4}$) | Mass | Node | Pressure ($10^{-4}$) | Mass |
|---|---|---|---|---|---|---|---|---|
| 4 | 0.1510 | 0.1377 | 15 | 0.9315 | 0.5219 | 26 | 0.0069 | 0.5753 |
| 5 | 0.0006 | 0.4647 | 19 | 0.0030 | 0.6039 | 29 | 0.0039 | 0.5154 |
| 8 | 0.0971 | 0.6426 | 20 | 0.0256 | 0.6601 | 30 | 0.0024 | 0.1810 |
| 10 | 0.0202 | 0.5502 | 22 | 0.0003 | 0.4405 | 31 | 0.0056 | 0.0197 |
| 11 | 0.0214 | 0.6647 | 24 | 0.0531 | 0.0282 | 34 | 0.0001 | 0.1180 |
| Bus | *e* | *f* | Bus | *e* | *f* | Bus | *e* | *f* |
| 2 | 0.0140 | 0.1432 | 21 | 0.0279 | 0.2822 | 33 | 0.1148 | 0.3191 |
| 4 | 0.0253 | 0.2825 | 22 | 0.0458 | 0.3155 | 34 | 0.1106 | 0.4260 |
| 6 | 0.0300 | 0.2845 | 25 | 0.0149 | 0.2061 | 35 | 0.0430 | 0.3209 |
| 9 | 0.0481 | 0.0504 | 28 | 0.0127 | 0.1786 | 36 | 0.0616 | 0.4832 |
| 12 | 0.0206 | 0.2661 | 30 | 0.0164 | 0.1725 | 37 | 0.0141 | 0.1701 |
| 15 | 0.0308 | 0.2596 | 31 | 0.0276 | 0.2113 | 38 | 0.0175 | 0.1878 |
| 18 | 0.0266 | 0.2449 | 32 | 0.0223 | 0.2444 | 39 | 0.0440 | 0.0618 |

TABLE II
TOTAL VARIANCES OF THE GAS PRESSURES UNDER WHITE GAUSSIAN NOISES

| Node | $\varepsilon_2(10^{-6})$ | Node | $\varepsilon_2(10^{-6})$ | Node | $\varepsilon_2(10^{-6})$ | Node | $\varepsilon_2(10^{-6})$ |
|---|---|---|---|---|---|---|---|
| 4 | 9.0401 | 12 | 0.8686 | 19 | 0.2369 | 28 | 0.0635 |
| 5 | 0.0495 | 13 | 0.0311 | 21 | 0.3192 | 30 | 0.1367 |
| 7 | 0.4586 | 15 | 9.4359 | 22 | 0.0200 | 31 | 0.3379 |
| 9 | 0.3916 | 17 | 0.1172 | 25 | 0.4688 | 33 | 0.0123 |
| 10 | 1.1353 | 18 | 0.0331 | 27 | 3.6086 | 34 | 0.0093 |

TABLE III
TOTAL VARIANCES OF MASS FLOWS AND VOLTAGES UNDER WHITE GAUSSIAN NOISES

| Pipe | $m_{ij}$ | $m_{ji}$ | Pipe | $m_{ij}$ | $m_{ji}$ | Pipe | $m_{ij}$ | $m_{ji}$ |
|---|---|---|---|---|---|---|---|---|
| 1,6 | 0.098 | 0.090 | 9,25 | 0.083 | 0.090 | 18,32 | 0.122 | 0.123 |
| 28,16 | 0.108 | 0.098 | 10,8 | 0.043 | 0.044 | 32,2 | 0.030 | 0.030 |
| 28,29 | 0.249 | 0.250 | 25,4 | 0.023 | 0.031 | 3,22 | 0.087 | 0.087 |
| 29,7 | 0.066 | 0.066 | 27,24 | 0.079 | 0.084 | 30,22 | 0.008 | 0.008 |
| 7,23 | 0.016 | 0.016 | 20,11 | 0.034 | 0.035 | 13,14 | 0.165 | 0.169 |
| 9,10 | 0.044 | 0.047 | 11,23 | 0.075 | 0.075 | 13,33 | 0.080 | 0.081 |
| Bus | $e\,(10^{-4})$ | $f(10^{-3})$ | Bus | $e\,(10^{-4})$ | $f(10^{-3})$ | Bus | $e\,(10^{-4})$ | $f(10^{-3})$ |
| 1 | 0.183 | 0.024 | 14 | 0.096 | 0.105 | 27 | 0.105 | 0.107 |
| 2 | 0.056 | 0.070 | 19 | 0.459 | 0.143 | 29 | 0.068 | 0.075 |
| 3 | 0.104 | 0.106 | 21 | 0.115 | 0.121 | 32 | 0.095 | 0.104 |
| 8 | 0.165 | 0.022 | 23 | 0.251 | 0.185 | 36 | 0.250 | 0.185 |
| 13 | 0.096 | 0.104 | 25 | 0.056 | 0.070 | 39 | 0.181 | 0.024 |

The estimating performance indexes are shown partly in Tab. V and VI. It is can be seen that filter coefficients are smaller than 1, meaning that the DSE method is effective. However, it should be noted that the values of bus 9 and 18 are larger than other values in Tab. V. The reason is that the fluctuations of power loads at bus 9 and 18 are stronger than others, as well as the voltages. So the predicting accuracies of the two buses are low, resulting in low estimating accuracies. As a result, the filter coefficients of the two buses are larger than others.

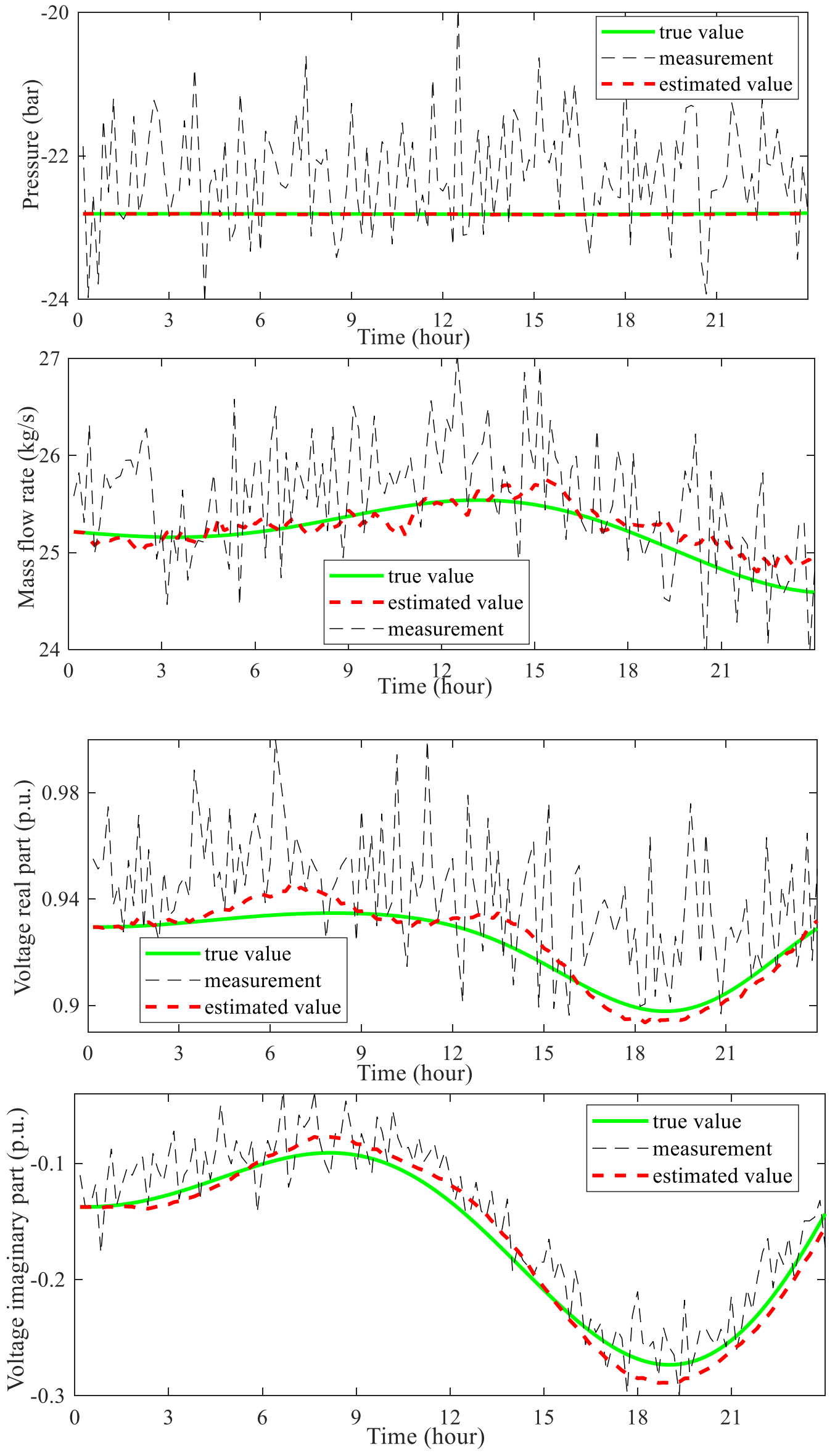

Fig. 3. Estimating results under non-zero mean noises.

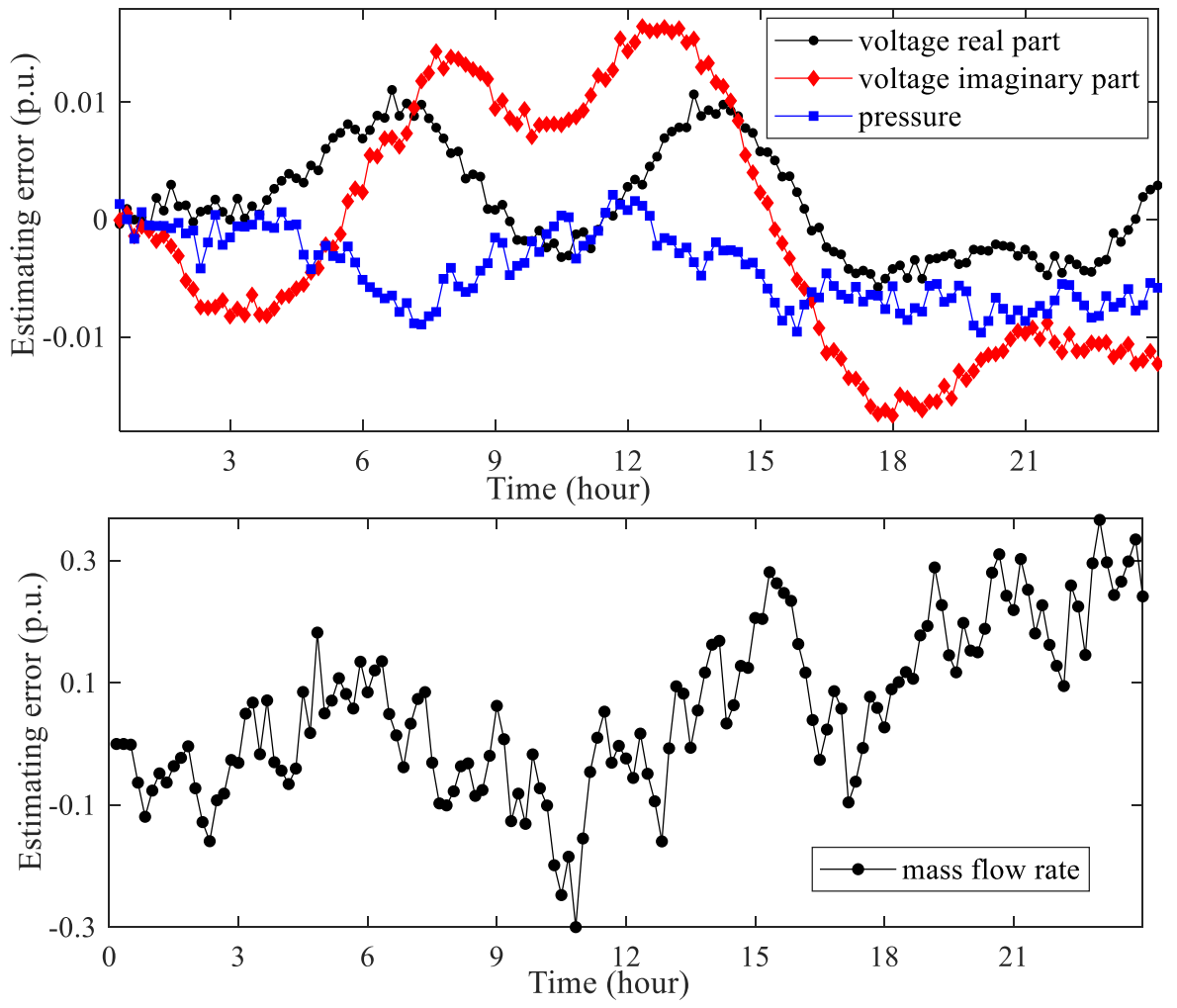

Fig. 4. Estimating errors under non-zero mean noises.

TABLE IV
THE VALUES OF INDEXES UNDER COLORED NOISES

| | | | | | | |
|---|---|---|---|---|---|---|
| Deviation (bar) | | | 0.2 | 0.4 | 0.6 | 0.8 |
| Node 24 | Pressure | $\varepsilon_1(10^{-4})$ | 0.0341 | 0.0200 | 0.0144 | 0.0130 |
| | | $\varepsilon_2(10^{-4})$ | 0.0603 | 0.1207 | 0.1408 | 0.1723 |
| Deviation (kg/s) | | | 0.2 | 0.6 | 1.0 | 1.4 |
| Node 4 | Mass flow rate | $\varepsilon_1$ | 0.0998 | 0.0890 | 0.0364 | 0.0253 |
| | | $\varepsilon_2$ | 0.0321 | 0.0490 | 0.0566 | 0.0613 |
| Deviation (p.u.) | | | 0.01 | 0.04 | 0.07 | 0.10 |
| Bus 12 | $e$ | $\varepsilon_1$ | 0.0175 | 0.0161 | 0.0118 | 0.0095 |
| | | $\varepsilon_2(10^{-4})$ | 0.1037 | 0.1685 | 0.5952 | 0.9377 |
| | $f$ | $\varepsilon_1$ | 0.2276 | 0.1847 | 0.0242 | 0.0136 |
| | | $\varepsilon_2(10^{-3})$ | 0.1120 | 0.1328 | 0.1453 | 0.2017 |

TABLE V
FILTER COEFFICIENTS UNDER CAUCHY DISTRIBUTION

| Node | Pressure ($10^{-6}$) | Mass | Node | Pressure ($10^{-6}$) | Mass | Node | Pressure ($10^{-6}$) | Mass |
|---|---|---|---|---|---|---|---|---|
| 4 | 5.3918 | 0.5078 | 15 | 6.0977 | 0.0001 | 26 | 0.0247 | 0.0006 |
| 10 | 0.4003 | 0.0068 | 22 | 0.0021 | 0.0084 | 31 | 0.0007 | 0.2049 |
| 11 | 0.0001 | 0.0008 | 24 | 36.8502 | 0.7486 | 34 | 0.0318 | 0.0075 |
| Bus | $e$ | $f$ | Bus | $e$ | $f$ | Bus | $e$ | $f$ |
| 2 | 0.0068 | 0.0234 | 21 | 0.0134 | 0.0543 | 33 | 0.1454 | 0.2184 |
| 9 | 0.6340 | 0.8751 | 28 | 0.0239 | 0.0261 | 36 | 0.0449 | 0.0659 |
| 18 | 0.2804 | 0.9389 | 32 | 0.2549 | 0.6043 | 39 | 0.0849 | 0.0555 |

TABLE VI
FILTER COEFFICIENTS UNDER LAPLACE DISTRIBUTION

| Node | Pressure ($10^{-4}$) | Mass ($10^{-3}$) | Node | Pressure ($10^{-4}$) | Mass ($10^{-3}$) | Node | Pressure ($10^{-4}$) | Mass ($10^{-3}$) |
|---|---|---|---|---|---|---|---|---|
| 4 | 0.1018 | 1.1010 | 15 | 0.2827 | 0.6608 | 26 | 0.0071 | 0.7620 |
| 8 | 0.0290 | 0.7884 | 20 | 0.0127 | 0.8155 | 30 | 0.0021 | 1.8839 |
| 11 | 0.0101 | 0.8544 | 24 | 0.1204 | 6.8644 | 34 | 0.0002 | 29.0139 |
| Bus | $e$ | $f$ | Bus | $e$ | $f$ | Bus | $e$ | $f$ |
| 2 | 0.0305 | 0.1959 | 21 | 0.0565 | 0.2316 | 33 | 0.2243 | 0.4886 |
| 9 | 0.0372 | 0.2083 | 28 | 0.0371 | 0.1856 | 36 | 0.1914 | 0.2545 |
| 18 | 0.0520 | 0.3286 | 32 | 0.0545 | 0.2896 | 39 | 0.0380 | 0.1717 |

## *E. Comparison with the Separated Method*

Based on the parameters in section *B*, the estimation total variances of the separated method are given in Tab. VII and VIII. It can be seen that most of the total variances of separated gas estimation are larger than the integrated method. The mean values of total variances of the separated gas pressure and mass flow estimation are $3.3775 \times 10^{-6}$ bar and 0.09997 kg/s, respectively, while the mean values of integrated method are $1.2995 \times 10^{-6}$ bar and 0.0831 kg/s ,respectively. It indicates that gas estimation accuracy of the integrated DSE is higher than the separated method.

## *F. Bad Data Condition*

To testify the robustness of the proposed method, bad data condition is considered on the following measurements: real and imaginary parts of voltage measurements of bus 32, gas pressure measurement of node 20 and mass flow rate measurement of gas load at node 4. The simulation results are shown in Fig. 5. It can be seen that the estimated values are not affected by the bad data. The reason is that the time-varying scalar can regulate the measurement variance matrix according to the measurement errors.

TABLE VII
THE GAS PRESSURES TOTAL VARIANCES OF SEPARATED DSEs

| Node | $\varepsilon_2(10^{-6})$ | Node | $\varepsilon_2(10^{-6})$ | Node | $\varepsilon_2(10^{-6})$ | Node | $\varepsilon_2(10^{-6})$ |
|---|---|---|---|---|---|---|---|
| 4 | 2.0791 | 12 | 0.7743 | 20 | 1.9436 | 28 | 0.0646 |
| 5 | 0.0575 | 13 | 0.0301 | 21 | 0.3639 | 29 | 0.2369 |
| 6 | 0.0407 | 14 | 0.1041 | 22 | 0.0227 | 30 | 0.0478 |
| 7 | 0.4603 | 15 | 71.3840 | 23 | 0.1506 | 31 | 0.2989 |
| 8 | 5.1470 | 16 | 0.1179 | 24 | 8.5858 | 32 | 0.0438 |
| 9 | 0.4440 | 17 | 0.1171 | 25 | 0.6147 | 33 | 0.0558 |
| 10 | 1.2077 | 18 | 0.0369 | 26 | 0.4386 | 34 | 0.0159 |
| 11 | 1.4525 | 19 | 0.2424 | 27 | 8.1240 | | |

TABLE VIII
THE MASS FLOW AND VOLTAGE TOTAL VARIANCES OF SEPARATED DSEs

| Pipe | $m_{ij}$ | $m_{ji}$ | Pipe | $m_{ij}$ | $m_{ji}$ | Pipe | $m_{ij}$ | $m_{ji}$ |
|---|---|---|---|---|---|---|---|---|
| 1,6 | 0.104 | 0.095 | 9,25 | 0.135 | 0.146 | 18,32 | 0.126 | 0.128 |
| 14,19 | 0.088 | 0.091 | 10,27 | 0.123 | 0.124 | 32,31 | 0.097 | 0.097 |
| 28,16 | 0.107 | 0.095 | 25,4 | 0.080 | 0.089 | 32,5 | 0.083 | 0.083 |
| 16,17 | 0.066 | 0.073 | 27,24 | 0.112 | 0.145 | 5,18 | 0.063 | 0.063 |
| 17,13 | 0.114 | 0.115 | 24,15 | 0.078 | 0.083 | 32,2 | 0.048 | 0.048 |
| 28,29 | 0.256 | 0.257 | 10,8 | 0.057 | 0.057 | 3,22 | 0.140 | 0.140 |
| 29,12 | 0.142 | 0.141 | 8,20 | 0.105 | 0.101 | 22,33 | 0.013 | 0.012 |
| 12,21 | 0.118 | 0.112 | 20,7 | 0.059 | 0.058 | 3,34 | 0.061 | 0.046 |
| 29,7 | 0.070 | 0.070 | 20,11 | 0.036 | 0.037 | 30,34 | 0.055 | 0.041 |
| 7,23 | 0.017 | 0.016 | 6,26 | 0.097 | 0.097 | 30,22 | 0.075 | 0.075 |
| 21,9 | 0.119 | 0.107 | 11,23 | 0.078 | 0.077 | 13,14 | 0.169 | 0.168 |
| 28,6 | 0.135 | 0.135 | 28,23 | 0.178 | 0.178 | 13,22 | 0.165 | 0.167 |
| 9,10 | 0.064 | 0.062 | 28,18 | 0.186 | 0.191 | 13,33 | 0.079 | 0.080 |

| Bus | $e\,(10^{-4})$ | $f(10^{-3})$ | Bus | $e\,(10^{-4})$ | $f(10^{-3})$ | Bus | $e\,(10^{-4})$ | $f(10^{-3})$ |
|---|---|---|---|---|---|---|---|---|
| 1 | 0.183 | 0.024 | 14 | 0.096 | 0.105 | 27 | 0.105 | 0.107 |
| 2 | 0.056 | 0.070 | 15 | 0.101 | 0.106 | 28 | 0.064 | 0.074 |
| 3 | 0.104 | 0.106 | 16 | 0.103 | 0.106 | 29 | 0.068 | 0.075 |
| 4 | 0.097 | 0.105 | 17 | 0.104 | 0.106 | 30 | 0.056 | 0.071 |
| 5 | 0.097 | 0.104 | 18 | 0.104 | 0.106 | 31 | 0.099 | 0.104 |
| 6 | 0.096 | 0.104 | 19 | 0.459 | 0.143 | 32 | 0.095 | 0.104 |
| 7 | 0.097 | 0.105 | 20 | 0.456 | 0.143 | 33 | 0.455 | 0.143 |
| 8 | 0.165 | 0.022 | 21 | 0.115 | 0.121 | 34 | 0.451 | 0.142 |
| 9 | 0.174 | 0.023 | 22 | 0.178 | 0.152 | 35 | 0.178 | 0.152 |
| 10 | 0.096 | 0.105 | 23 | 0.251 | 0.185 | 36 | 0.250 | 0.185 |
| 11 | 0.096 | 0.104 | 24 | 0.263 | 0.186 | 37 | 0.057 | 0.070 |
| 12 | 0.097 | 0.106 | 25 | 0.056 | 0.070 | 38 | 0.068 | 0.075 |
| 13 | 0.096 | 0.104 | 26 | 0.061 | 0.070 | 39 | 0.181 | 0.024 |

The simulations are carried out on the computer with 1.60GHz CPU, and the average time consumption of one step is 362.5ms for the integrated DSEs, which can meet the real-time requirements in the IGES of the same scale. Practically, the calculating step is several minutes, and the computing capability can be much stronger, so the method can satisfy the real-time requirements in practical system.

## VI. CONCLUSION

In this paper, a robust DSE method for IGESs is proposed, which is applied on an IGES with Gaslib-40 and IEEE 39-bus system coupled by five GTUs, under different measurement conditions. The filter coefficient and total variance are used to evaluate the performances of the DSE method. The results show that all of the filter coefficients are smaller than 1 under the white Gaussian noises condition, meaning that the DSE is effective. Furthermore, the method is studied under the non-zero mean measurement noises and non-Gaussian noises conditions. The results show that the method is effective. The integrated DSE are compared with separated method, and the results show that the gas estimation accuracy of integrated DSE is higher. Additionally, the proposed method is applied under bad data condition, and the estimation results indicate that the proposed robust DSE still shows a good performance.

Further works can be carried out in the following areas: the accurate calculating method of predicting errors; the robust DSE algorithm against bad data, colored noises and model parameter errors. It is also interesting to extend the proposed dynamic state estimation methodology for developing a distributed state estimator for integrated energy systems [34, 35]. Another interesting topic is to design a state estimator of power systems by using advanced machine learning techniques, such as deep learning [36] or reinforcement learning [37].

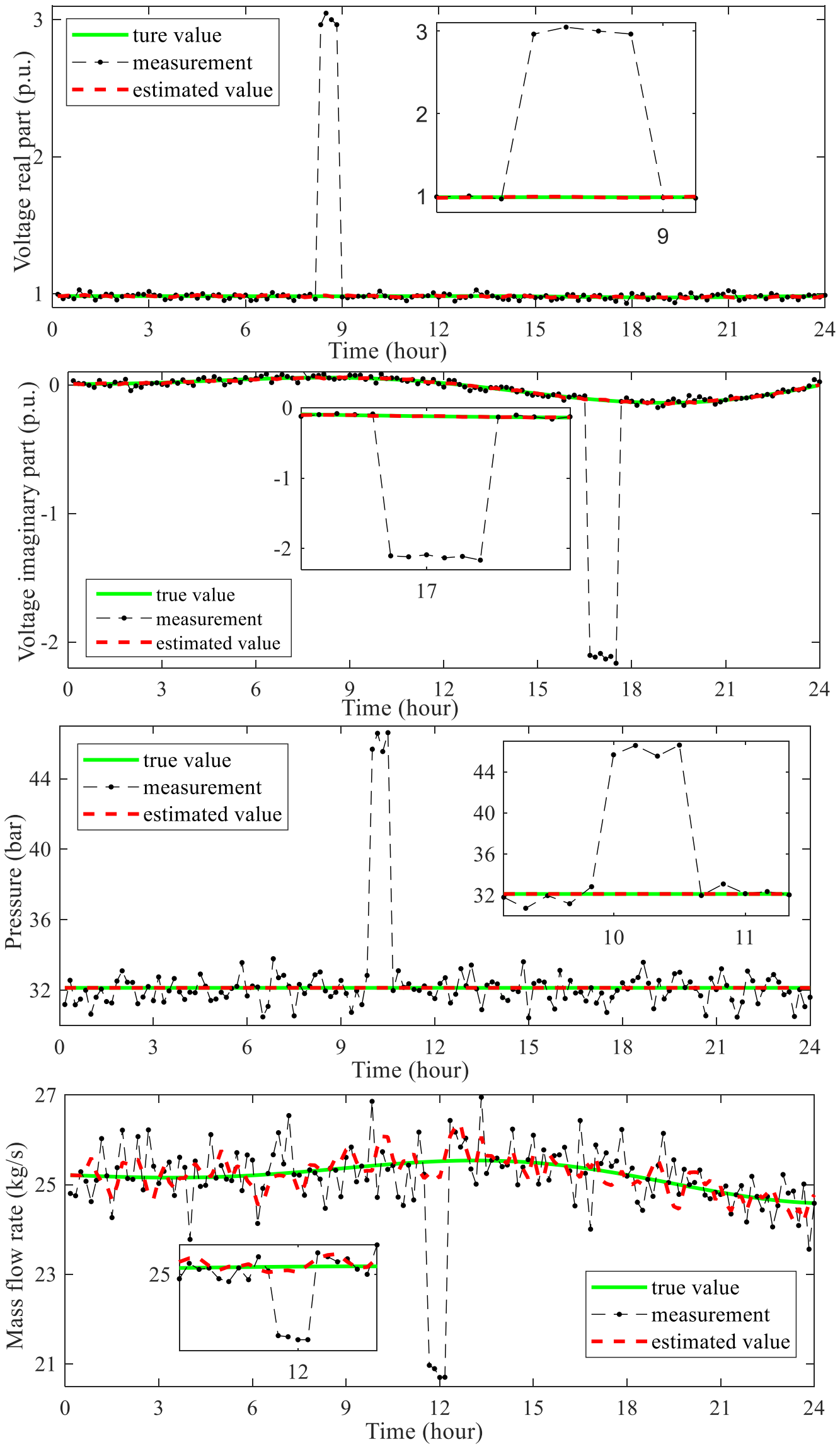

Fig. 5. Estimating results under bad data conditions.

## APPENDIX A

The details of power system measurement matrix are:

$$\begin{cases} h_{\mathrm{V}}(2b_{\mathrm{M}}-1,2i-1)=h_{\mathrm{V}}(2b_{\mathrm{M}},2i)=1, & \text{if } b_{\mathrm{M}} \triangleright i \\ h_{\mathrm{V}}=0, & \text{eles} \end{cases} \tag{38}$$

$$\boldsymbol{H}_{\mathrm{IB}}=\begin{matrix} \\ 2l_{\mathrm{B}}-1 \\ 2l_{\mathrm{B}} \\ \\ \end{matrix}\overset{\begin{matrix} \quad 2i-1 & \quad\quad 2i & \quad 2j-1 & \quad 2j \end{matrix}}{\begin{bmatrix} & \vdots & \cdots & \vdots & \cdots & \vdots & \cdots & \vdots & \\ \cdots & g_{ij}+g_{i0} & \cdots & -b_{ij}-b_{i0} & \cdots & -g_{ij} & \cdots & b_{ij} & \cdots \\ \cdots & b_{ij}+b_{i0} & \cdots & g_{ij}+g_{i0} & \cdots & -b_{ij} & \cdots & -g_{ij} & \cdots \\ & \vdots & \cdots & \vdots & \cdots & \vdots & \cdots & \vdots & \end{bmatrix}} \tag{39}$$

$$\begin{cases} h_{\mathrm{IN}}(2b_{\mathrm{M}}-1,2j-1)=h_{\mathrm{IN}}(2b_{\mathrm{M}},2j)=G_{ij} \\ h_{\mathrm{IN}}(2b_{\mathrm{M}},2j-1)=-h_{\mathrm{IN}}(2b_{\mathrm{M}}-1,2j)=B_{ij} \end{cases}, j \in \mathbb{R}^{n_{\mathrm{B}}} \tag{40}$$

where, $h_{\mathrm{V}}$ and $h_{\mathrm{IN}}$ are the element of $\boldsymbol{H}_{\mathrm{V}}$ and $\boldsymbol{H}_{\mathrm{IN}}$, respectively; $b_{\mathrm{M}} \triangleright i$ means the $b_{\mathrm{M}}^{\mathrm{th}}$ voltage phasor measurements is the voltage of bus $i$, $b_{\mathrm{M}} \in \mathbb{R}^{n_{\mathrm{ZB}}}$; $g_{ij}$ and $b_{ij}$ are the conductance and susceptance of branch $ij$ respectively; $g_{i0}$ and $b_{i0}$ are the shunt conductance and susceptance of bus $i$ respectively, $l_{\mathrm{B}}$ is the branch number, $l_{\mathrm{B}} \in \mathbb{R}^{n_{\mathrm{ZC}}}$; $(i,j)$ represents the $i^{\mathrm{th}}$ row and the $j^{\mathrm{th}}$ column; $G_{ij}$ and $B_{ij}$ are the real and imaginary parts of the $i^{\mathrm{th}}$ row $j^{\mathrm{th}}$ column element in the node admittance matrix, respectively; $n_{\mathrm{B}}$ is the number of buses.

## APPENDIX B

The details of system matrixes $\boldsymbol{A}_{11}$, $\boldsymbol{A}_{12}$, $\boldsymbol{A}_{21}$ and $\boldsymbol{A}_{22}$ are :

$$\boldsymbol{A}_{11}(l,i)=\begin{cases} c_{ij}^{\mathrm{r}}, & \text{if } i \propto l,\ j \propto l \\ 0, & \text{else} \end{cases}, \boldsymbol{A}_{11} \in \mathbb{R}^{n_{\mathrm{P}} \times n_{\mathrm{N}}} \tag{41}$$

$$\boldsymbol{A}_{12}=\begin{bmatrix} -\xi_1 & \xi_1 & & & & \mathbf{0} \\ & & -\xi_2 & \xi_2 & & \\ & & & \ddots & & \\ \mathbf{0} & & & & -\xi_{n_{\mathrm{L}}} & \xi_{n_{\mathrm{L}}} \end{bmatrix}, \boldsymbol{A}_{12} \in \mathbb{R}^{n_{\mathrm{P}} \times 2n_{\mathrm{P}}} \tag{42}$$

$$\xi_l=\frac{\Delta t}{L_{ij}a_{ij}}, i \propto l, j \propto l \tag{43}$$

$$\boldsymbol{A}_{21}(l,i)=\begin{cases} -c_{ij}^{\mathrm{r}}\beta_l, & \text{if } i \propto l, j \propto l, i<j \\ c_{ji}^{\mathrm{r}}\beta_l, & \text{if } i \propto l, j \propto l, i>j \\ 0, & \text{else} \end{cases}, \boldsymbol{A}_{21} \in \mathbb{R}^{n_{\mathrm{P}} \times n_{\mathrm{N}}} \tag{44}$$

$$\beta_l=\frac{a_{ij}\Delta t c_{\mathrm{S}}}{L_{ij}}, i \propto l, j \propto l \tag{45}$$

$$\boldsymbol{A}_{22}(l,i)=\begin{cases} \gamma_l-1, & \text{if } i \propto l, j \propto l, i<j \\ \gamma_l+1, & \text{if } i \propto l, j \propto l, i>j \\ 0, & \text{else} \end{cases}, \boldsymbol{A}_{22} \in \mathbb{R}^{n_{\mathrm{P}} \times 2n_{\mathrm{P}}} \tag{46}$$

$$\gamma_l=\frac{f\,|\bar{v}_{\mathrm{G}}|\,\Delta t}{4d_{ij}a_{ij}}, i \propto l, j \propto l \tag{47}$$

where, $i \propto l$ means node $i$ is one end of pipeline $l$.

## APPENDIX C

The details of boundary condition matrixes $\boldsymbol{B}_{11}$ and $\boldsymbol{B}_{22}$ are :

$$\boldsymbol{B}_{11}(s,i)=\begin{cases} 1, & i: \text{source node} \\ 0, & i: \text{sink node} \end{cases}, \boldsymbol{B}_{11} \in \mathbb{R}^{n_{\mathrm{S}} \times n_{\mathrm{N}}} \tag{48}$$

$$\boldsymbol{B}_{22}(j,2l)=\begin{cases} 1, & j: \text{sink node}, j \propto l \\ 0, & \text{else} \end{cases}, \boldsymbol{B}_{22} \in \mathbb{R}^{n_{\mathrm{SI}} \times 2n_{\mathrm{P}}} \tag{49}$$

$$\boldsymbol{B}_{22}(i,2l-1)=\begin{cases} -1, & i: \text{sink node}, i \propto l \\ 0, & \text{else} \end{cases} \tag{50}$$

$$\boldsymbol{B}_{33}(v,2l_{\mathrm{V}})=\begin{cases} 1, & v: \text{virtual node}, v \propto l_{\mathrm{V}} \\ 0, & \text{else} \end{cases} \tag{51}$$

$$\boldsymbol{B}_{33}(v,2l_{\mathrm{V}}-1)=\begin{cases} -1, & v: \text{virtual node}, v \propto l_{\mathrm{V}} \\ 0, & \text{else} \end{cases} \tag{52}$$

## APPENDIX D

The details of gas system measurement matrixes $\boldsymbol{H}_{\mathrm{G}}$ are:

$$\boldsymbol{H}_{\mathrm{G}}=\begin{bmatrix} \boldsymbol{H}_{\mathrm{P}} & \mathbf{0}_{n_{\mathrm{ZP}} \times 2n_{\mathrm{P}}} \\ \mathbf{0}_{n_{\mathrm{ZF}} \times n_{\mathrm{N}}} & \boldsymbol{H}' \end{bmatrix} \tag{53}$$

$$\begin{cases} \boldsymbol{H}_{\mathrm{P}}(i,j)=c_{\mathrm{S}}, & \text{if } i \triangleq j \\ \boldsymbol{H}_{\mathrm{P}}(i,j)=0, & \text{else} \end{cases}, \boldsymbol{H}_{\mathrm{P}} \in \mathbb{R}^{n_{\mathrm{ZP}} \times n_{\mathrm{N}}} \tag{54}$$

$$\begin{cases} \boldsymbol{H}'(j,2l)=1 \\ \boldsymbol{H}'(i,2l-1)=-1 \end{cases} i,j \propto l,\ i<j,\ \boldsymbol{H}' \in \mathbb{R}^{n_{\mathrm{N}} \times 2n_{\mathrm{P}}} \tag{55}$$

where, $n_{\mathrm{ZP}}$ and $n_{\mathrm{ZF}}$ are the numbers of pressure and load flow measurements, respectively; $i \triangleq j$ represents the $i^{\mathrm{th}}$ pressure measurement is the pressure of node $j$.